\documentclass{article}
\usepackage{arxiv}

\usepackage{graphicx}
\usepackage{multirow}
\usepackage{amsmath,amssymb,amsfonts}
\usepackage{amsthm}
\usepackage{mathrsfs}
\usepackage[title]{appendix}
\usepackage{textcomp}
\usepackage{booktabs}
\usepackage{algorithm}
\usepackage{algorithmicx}
\usepackage{algpseudocode}
\usepackage{listings}
\usepackage[table]{xcolor}
\usepackage{tabularx}
\usepackage{adjustbox}
\usepackage{siunitx}
\usepackage{makecell}
\usepackage{tikz}
\usetikzlibrary{arrows.meta,positioning}
\usepackage{float}
\usepackage[hidelinks]{hyperref}
\hypersetup{
	colorlinks,
	linkcolor={red!50!black},
	citecolor={blue!50!black},
	urlcolor={green!40!black}
}
\usepackage{url}
\usepackage{natbib}
\title{MyGardenBird: A Machine-Learning-Ready Bird Sound Dataset for Twelve Common Malaysian Birds}

\author{%
  Muhammad Mun'im Ahmad Zabidi$^{1,2}$ \quad
  Mohd Yamani Idna Idris$^{1}$ \quad
  Norisma Idris$^{1}$ \\[4pt]
  $^{1}$Faculty of Computer Science and Information Technology, Universiti Malaya,\\
  50603 Kuala Lumpur, Malaysia \\
  $^{2}$Faculty of Electrical Engineering, Universiti Teknologi Malaysia,\\
  81310 Johor Bahru, Johor, Malaysia \\[4pt]
  \texttt{yamani@um.edu.my} \quad (corresponding author)
}

\begin{document}

\maketitle

\begin{abstract}
Bioacoustic datasets from tropical regions remain limited, in part due to the absence of reproducible workflows for aggregating recordings from public archives. We present \textbf{MyGardenBird}, a curated dataset of bird vocalisations representing twelve common species across Peninsular Malaysia and the Indo-Malayan region. Recordings were sourced from Xeno-canto and processed through species-level filtering, manual spectrogram segmentation, and quality control checks. The primary release comprises 7,200 manually validated audio clips (16 kHz, 16-bit PCM mono WAV), balanced at 600 three‑second clips per species (6.0 hours total) derived from 1,381 distinct recordings. Metadata includes geospatial coordinates, vocalisation categories, and signal‑to‑noise ratio (SNR) values (range: 0.83–59.18 dB; mean: 15.80 dB). A supplementary 44.1 kHz version is also provided. To mitigate data leakage, dataset partitions are defined at the source‑recording level. Baseline classification experiments using convolutional neural networks on Mel‑spectrograms achieved test accuracies of 92–96\%, indicating strong interspecies separability. Limitations include reliance on single‑annotator curation; however, validation with BirdNET confirmed label consistency. MyGardenBird is openly available at \url{https://doi.org/10.5281/zenodo.20306877} under a CC BY‑NC‑SA 4.0 licence. Complete preprocessing code accompanies the release to support reproducibility and future expansion.
\end{abstract}



\section{Background and Summary}\label{sec1}

Birds are widely regarded as sensitive indicators of ecosystem health and environmental change \citep{fraixedas2020state}. Consequently, Passive Acoustic Monitoring (PAM) has emerged as an effective and non-invasive approach that complements conventional field surveys by enabling large-scale and long-term monitoring of avian populations \citep{funosas2026global}. The effectiveness of PAM-based systems, however, depends critically on the availability of well-curated and accurately labelled training datasets. Although public repositories such as \textit{Xeno-canto} and \textit{iNaturalist}, together with benchmark initiatives such as BirdCLEF \citep{kahl2023overview}, have substantially expanded access to avian acoustic recordings, the global bioacoustic data landscape remains unevenly represented. Existing curated datasets are disproportionately concentrated in temperate regions of Europe and North America \citep{perez2023birdnet}, while many recordings exhibit substantial variation in duration and are weakly labelled, necessitating extensive preprocessing before they can be effectively employed for machine learning applications \citep{sebastian2025geographic}.

These geographic and curation biases severely bottleneck classifier performance in the field. A recent global evaluation revealed that while BirdNET achieved PR AUC scores of 0.16–0.23 in North America and Europe, its performance plummeted to just 0.03–0.04 in Africa and Asia when tested on complex, species-rich soundscapes \citep{funosas2026global}. The study traced this drop-off directly to a lack of training data coverage rather than a flaw in model architecture. This shortfall is especially acute in Southeast Asia—a critical biodiversity hotspot. Malaysia alone boasts over 850 bird species \citep{mybis2026,lepage2026}, yet pre-segmented, quality-controlled datasets for training and evaluation are still incredibly rare in the region.

To address this limitation, we developed the \textbf{MyGardenBird} dataset, a curated bioacoustic corpus tailored for machine learning applications. Unlike conventional audio repositories that primarily serve archival purposes, from which this dataset is derived, {MyGardenBird} provides standardised three‑second clips, balanced class distributions, signal‑quality metadata, and predefined training, validation, and test partitions. The initial release comprises 7,200 manually reviewed clips evenly distributed across 12 common species (600 clips per species; 6.0\,h total). Recordings were sourced from \textit{Xeno-canto} and processed through a six-step curation pipeline. All files are provided in 16‑bit PCM mono WAV format at 16\,kHz, with an additional subset preserved at the original 44.1\,kHz sampling rate for applications requiring higher fidelity.

To prevent data leakage, we partitioned the train, validation, and test splits at the source-recording level using mixed-integer programming. We then checked our label consistency using BirdNET\,v2.4 in a training-free setup. It achieved a striking 97.94\% accuracy at 16\,kHz across all 7,200 clips, and 98.06\% accuracy at 44.1\,kHz across 6,950 clips. These high marks suggest that our clips genuinely represent their target species and are free from systematic mislabelling relative to the broader Xeno-canto database. To establish a baseline, we ran classification experiments using three different CNN architectures; fed with Mel-spectrogram features, they achieved a stable 92--96\% accuracy.

The complete dataset, metadata, and curation code are publicly available at \url{https://doi.org/10.5281/zenodo.20306877} under a CC~BY-NC-SA~4.0 licence. While the core dataset is intentionally limited to twelve species, annotated by a single curator, and focused on clearly identifiable vocalisations, we have provided an avenue for expansion. Specifically, we packaged supplementary recordings for the Common Myna and Zebra Dove into a separate `plus' zip archive within the Zenodo repository. These species were excluded from the regional core because ASEAN/Indo-Malayan sources yielded only 75 and 577 usable clips, respectively; these `plus' archives supplement those counts with global Xeno-canto recordings for users who do not require strict regional provenance. Looking ahead, we plan to scale up species coverage and integrate recordings from additional citizen-science platforms, applying these same rigorous standards for metadata quality, geographic origin, and licensing.

\section{Methods}\label{sec2}

\begin{figure}[!htbp]
\centering
\includegraphics{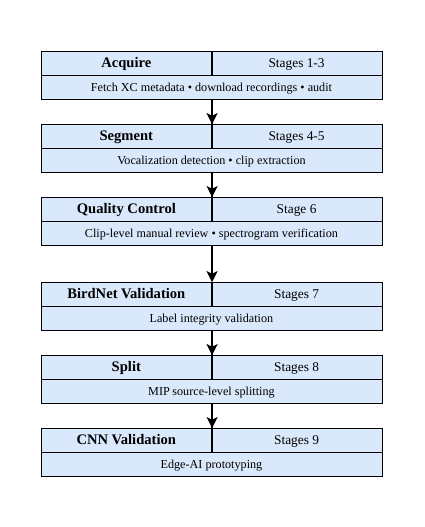}
\caption{MyGardenBird curation pipeline, comprising six sequential steps implemented across nine Python scripts (Stage~1 through Stage~9; source code at \url{https://github.com/mun3im/MyGardenBird}). Each \emph{step} denotes a logical pipeline phase; each \emph{Stage} denotes a numbered Python script. The Segment and Quality Control steps encompass multiple Stages. BirdNET zero-shot validation (Stage~7) provides a same-source label-consistency check on all curated clips before partitioning. Mixed-integer programming (MIP) is adopted as the default splitting algorithm in Stage~8. Stage~9 benchmarks CNN classifiers as a prototype for embedded deployment.}
\label{fig:pipeline}
\end{figure}

Figure~\ref{fig:pipeline} outlines the six-step curation pipeline; each step is described in detail in the subsections below.

\subsection{Species Selection and Dataset Naming}

\begin{table}[t]
	\centering\footnotesize
	\setlength{\tabcolsep}{5pt}
	\caption{Two-stage species-selection funnel applied to all 50 \textit{MY Garden Birdwatch} survey species~\citep{puan2019influence}. Selected species (upper block) and excluded species with explicit justifications (lower block).}
	\label{tab:shortlist}
	\begin{tabularx}{\textwidth}{@{}>{\raggedright\arraybackslash}p{0.25\textwidth}rrr>{\raggedright\arraybackslash}p{0.4\textwidth}@{}}
		\toprule
		\textbf{Species} & \makecell[r]{\textbf{Regional}\\\textbf{hours}} & \makecell[r]{\textbf{Regional}\\\textbf{files}} & \makecell[r]{\textbf{Grade}\\\textbf{A/B}} & \textbf{Exclusion reason} \\
		\midrule
		\multicolumn{5}{@{}l}{\textit{Selected (12 species)}} \\[2pt]
		Asian Koel              & 1.80 & 215 & 189 & --- \\
		Collared Kingfisher     & 1.09 & 147 & 123 & --- \\
		Common Iora             & 2.45 & 233 & 131 & --- \\
		Common Tailorbird       & 2.33 & 250 & 175 & --- \\
		Coppersmith Barbet      & 1.25 & 136 &  51 & --- \\
		Large-tailed Nightjar   & 2.98 & 207 & 143 & --- \\
		Olive-backed Sunbird    & 4.30 & 340 & 246 & --- \\
		Pied Fantail            & 1.32 & 142 & 124 & --- \\
		Spotted Dove            & 1.41 & 127 &  82 & --- \\
		White-breasted Waterhen & 1.63 & 137 & 107 & --- \\
		White-throated Kingfisher & 1.02 & 122 &  95 & --- \\
		Yellow-vented Bulbul    & 1.76 & 136 &  77 & --- \\
		\midrule
		\multicolumn{5}{@{}l}{\textit{Excluded (7 species)}} \\[2pt]
		Greater Racket-tailed Drongo  & 5.12 & 408 & 335 & Restricted to deep forest/jungle; prolific vocal mimic introducing extreme intra-class variability. \\
		Brown-throated Sunbird     & 1.82 & 178 & 125 & Congener of Olive-backed Sunbird; excluded to maximise inter-class acoustic separability. \\
		Ashy Tailorbird            & 1.78 & 183 & 153 & Congener of Common Tailorbird; excluded to maximise inter-class acoustic separability. \\
		Lineated Barbet            & 1.38 & 158 & 108 & Congener of Coppersmith Barbet; excluded to maximise inter-class acoustic separability. \\
		Zebra Dove                 & 1.14 & 131 &  93 & Insufficient localized recordings; yielded only 577 clips within target regional boundaries. \\
		Scarlet-backed Flowerpecker & 1.02 & 125 &  65 & Omitted from basic MNS guide; excluded from primary ingest pipeline. \\
		Common Myna$^\dagger$       & 0.60 &  57 &  45 & Insufficient localized recordings; yielded only 75 clips within target regional boundaries. \\
		\bottomrule
	\end{tabularx}
\end{table}

We based our species selection on the MY Garden Birdwatch initiative, an annual citizen-science survey organised by the Malaysian Nature Society (MNS). The programme monitors bird populations observed during 30-minute survey periods in residential areas and urban green spaces, supported by an identification guide that highlights approximately 30 common species from a broader pool of 50 surveyed species \citep{puan2019influence}. We adopted this 50-species pool as our initial sampling frame and named the dataset MyGardenBird to reflect its connection to, and recognition of, this well-established citizen-science initiative.

To build a reliable corpus for machine learning, we narrowed this 50-species pool down using a two-step screening process (summarised in Table~\ref{tab:shortlist}). First, we combed the Xeno-canto archive for species with at least 1.0~h of audio recorded within the ASEAN and Indo-Malayan regions. This initial regional sweep yielded 18 candidate species. We also provisionally kept ($\dagger$) the Common Myna (\textit{Acridotheres tristis}) as a 19th candidate; even though it fell below our raw audio volume threshold, it is simply too ubiquitous in Malaysian urban environments to leave out.

In the second step, we evaluated these 19 candidates against four core criteria: (1) the availability of high-quality recordings under permissive Creative Commons licences; (2) clear acoustic distinctiveness and inter-class separability; (3) regional abundance across Peninsular Malaysia; and (4) balanced representation across urban, peri-urban, and forest-edge habitats.

Ten species cleared all four hurdles. To round out the dataset to a balanced dozen, we pulled the Collared Kingfisher (\textit{Todiramphus chloris}) and Large-tailed Nightjar (\textit{Caprimulgus macrurus}) from the broader survey frame. While neither is featured in the simplified MNS basic identification guide, both are deeply characteristic of Malaysian peri-urban habitats, are acoustically unmistakable, and offered a wealth of high-quality regional recordings. This brought our final cohort to twelve evenly balanced species.

\begin{table}[t]
	\centering
	\footnotesize
	\caption{Composition of the MyGardenBird dataset, detailing class balances across audio sampling rates and the unique Xeno-canto (XC) source files utilised.}
	\label{tab:species}
	\begin{tabular*}{\textwidth}{@{\extracolsep{\fill}}cllcccc@{}}
		\toprule
		\textbf{No.} & \textbf{Common Name} & \textbf{Scientific Name} & \makecell{\textbf{eBird}\\\textbf{Code}} & \makecell{\textbf{16\,kHz}\\\textbf{Clips}} & \makecell{\textbf{44.1\,kHz}\\\textbf{Clips}} & \makecell{\textbf{XC}\\\textbf{Files}} \\
		\midrule
		1  & Asian Koel                & \textit{Eudynamys scolopaceus}      & asikoe2 & 600 & 574 & 136 \\
		2  & Collared Kingfisher        & \textit{Todiramphus chloris}        & colkin1 & 600 & 571 & 118 \\
		3  & Common Iora                & \textit{Aegithina tiphia}           & comior1 & 600 & 598 & 105 \\
		4  & Common Tailorbird          & \textit{Orthotomus sutorius}        & comtai1 & 600 & 577 & 96  \\
		5  & Coppersmith Barbet         & \textit{Psilopogon haemacephalus}   & copbar1 & 600 & 574 & 120 \\
		6  & Large-tailed Nightjar      & \textit{Caprimulgus macrurus}       & latnig2 & 600 & 579 & 108 \\
		7  & Olive-backed Sunbird       & \textit{Cinnyris jugularis}         & olbsun4 & 600 & 578 & 105 \\
		8  & Spotted Dove               & \textit{Spilopelia chinensis}       & spodov  & 600 & 599 & 97  \\
		9  & White-breasted Waterhen    & \textit{Amaurornis phoenicurus}     & whbwat1 & 600 & 572 & 115 \\
		10 & White-throated Kingfisher  & \textit{Halcyon smyrnensis}         & whtkin2 & 600 & 570 & 123 \\
		11 & Yellow-vented Bulbul       & \textit{Pycnonotus goiavier}        & yevbul1 & 600 & 581 & 118 \\
		12 & Pied Fantail               & \textit{Rhipidura javanica}         & piefan1 & 600 & 577 & 140 \\
		\midrule
		& \textbf{Total}             &                                     &         & \textbf{7,200} & \textbf{6,950} & \textbf{1,381} \\
		\bottomrule
	\end{tabular*}
\end{table}

During the pipeline's Quality Control (QC) step, we conducted a thorough manual review of both the acoustic structure and overall quality of the clips, systematically filtering out any recordings with overlapping species vocalisations or corrupted audio. This rigorous scrubbing left us with a final 16\,kHz master dataset of 7,200 fixed 3.0-second clips drawn from 1,381 unique source recordings. The dataset is perfectly balanced, offering 600 clips per class across the 12 selected species (Table~\ref{tab:species}). We have also provided a matching 44.1\,kHz subset for applications that benefit from higher spectral resolution.

Our 600-clip target was not just an arbitrarily chosen round number; rather, it represents the absolute practical upper limit of what we could achieve under strict ASEAN/Indo-Malayan regional constraints and our high quality standards.

Two clear observations bear this out. First, the Zebra Dove had to be excluded from the core dataset entirely because we only managed to extract 577 usable clips, even after exhausting every single regional recording available on Xeno-canto. Relaxing our quality filters simply did not close the gap. Second, both the Pied Fantail and Yellow-vented Bulbul barely hit the 600-clip mark, and doing so required us to include Grade~D recordings, which is the lowest quality tier permitted by our pipeline. Pushing our target any higher would have forced us to either severely compromise on audio quality or abandon our balanced class design.

Given these ecological and archiving constraints, 600 clips per species sits comfortably within the 50--1,000 clips-per-class range typically seen in major bioacoustic benchmarks \citep{kahl2021birdnet,stowell2022computational}. We found this volume to be a sweet spot: it is substantial enough to capture rich within-species vocal variations, yet compact enough to keep training and deployment practical for resource-constrained TinyML and edge hardware.

\subsection{Source Data Acquisition}

We sourced all our raw recordings from the Xeno-canto archive, selecting only those uploaded under Creative Commons licences that permit non-commercial redistribution. While all twelve target species are native to Peninsular Malaysia and the wider Indo-Malayan biogeographic region, we expanded our geographic scope to neighbouring countries when local recordings alone fell short of our 600-clip target. Specifically, we restricted our downloads to the ASEAN region, bounding our search to longitudinal coordinates between $60^\circ$ and $140^\circ$.

To filter this initial pool, we evaluated recordings against four core criteria: (i) a recorder-assigned quality rating of A–C, though we selectively incorporated Grade~D files if higher-grade regional sources were completely exhausted (this was necessary for the Pied Fantail and Yellow-vented Bulbul); (ii) unambiguous species identification; (iii) minimal overlapping vocalisations; and (iv) sufficient acoustic bandwidth for precise segmentation. To preserve raw audio integrity from the very start of our workflow, we immediately converted all downloaded MP3 files into a lossless, monaural FLAC format.

\begin{figure}[!htbp]
	\centering
	\includegraphics[width=\textwidth]{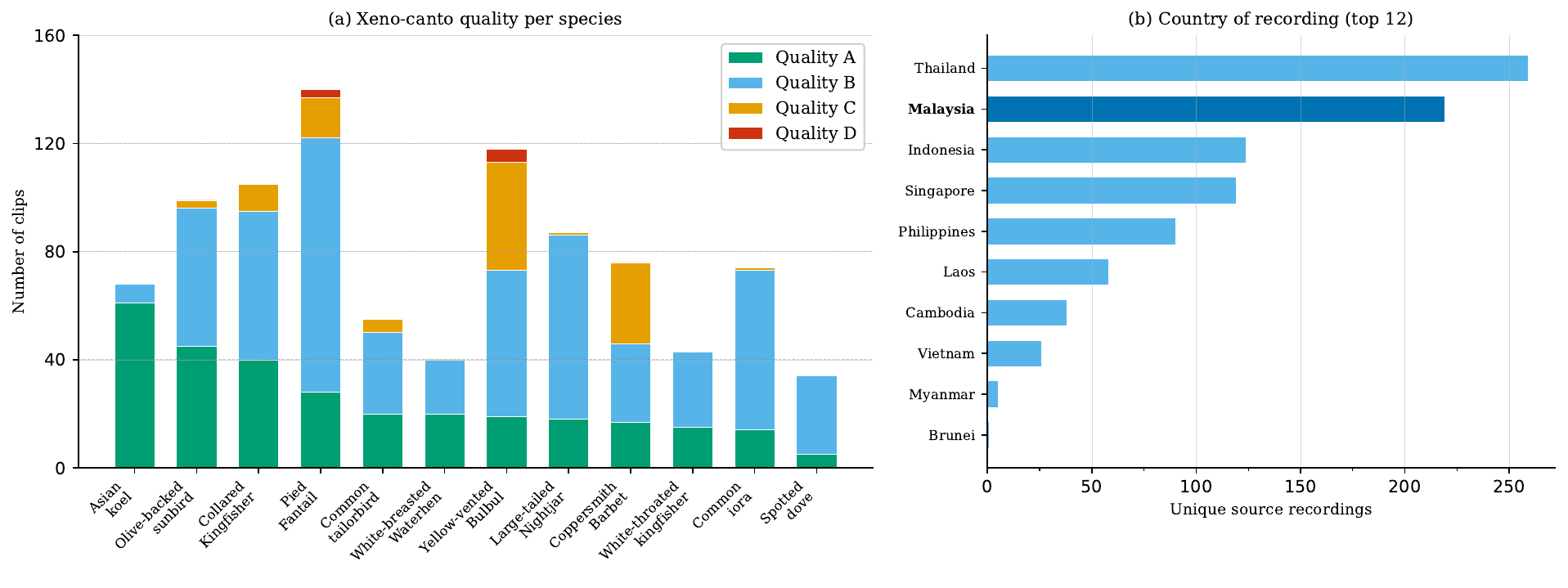}
	\caption{(a) Xeno-canto quality grade composition per species. Grade~A recordings were prioritised, with Grades~B and~C incorporated as necessary. Grade~D recordings were used only for Pied Fantail (3 sources) and Yellow-vented Bulbul (5 sources), where higher-grade regional sources were insufficient to reach the 600-clip target. (b) Country of origin for the 1,381 unique source recordings, derived from contributor metadata. While Malaysia remains the primary source, the broader distribution reflects the shared Indo-Malayan range of the twelve target species.}
	\label{fig:quality-country}
\end{figure}

Rather than artificially sanitizing the data, our dataset deliberately captures the wide range of equipment quality inherent to citizen-science archives, spanning everything from professional parabolic microphones to consumer smartphones. This real-world heterogeneity is reflected in our clip-level SNR distribution, which ranges from $0.83$ to $59.18$\,dB (Figure~\ref{fig:snr}). By leaving this variation intact, we ensure that models trained on MyGardenBird are inherently robust to the diverse acoustic conditions and hardware configurations they will encounter in the field. Figure~\ref{fig:quality-country} breaks down the distribution of these quality grades by species and illustrates the geographic origins of our source recordings.

\subsection{Segmentation Workflow}

\begin{figure}[!htbp]
	\centering
	\includegraphics[width=\textwidth]{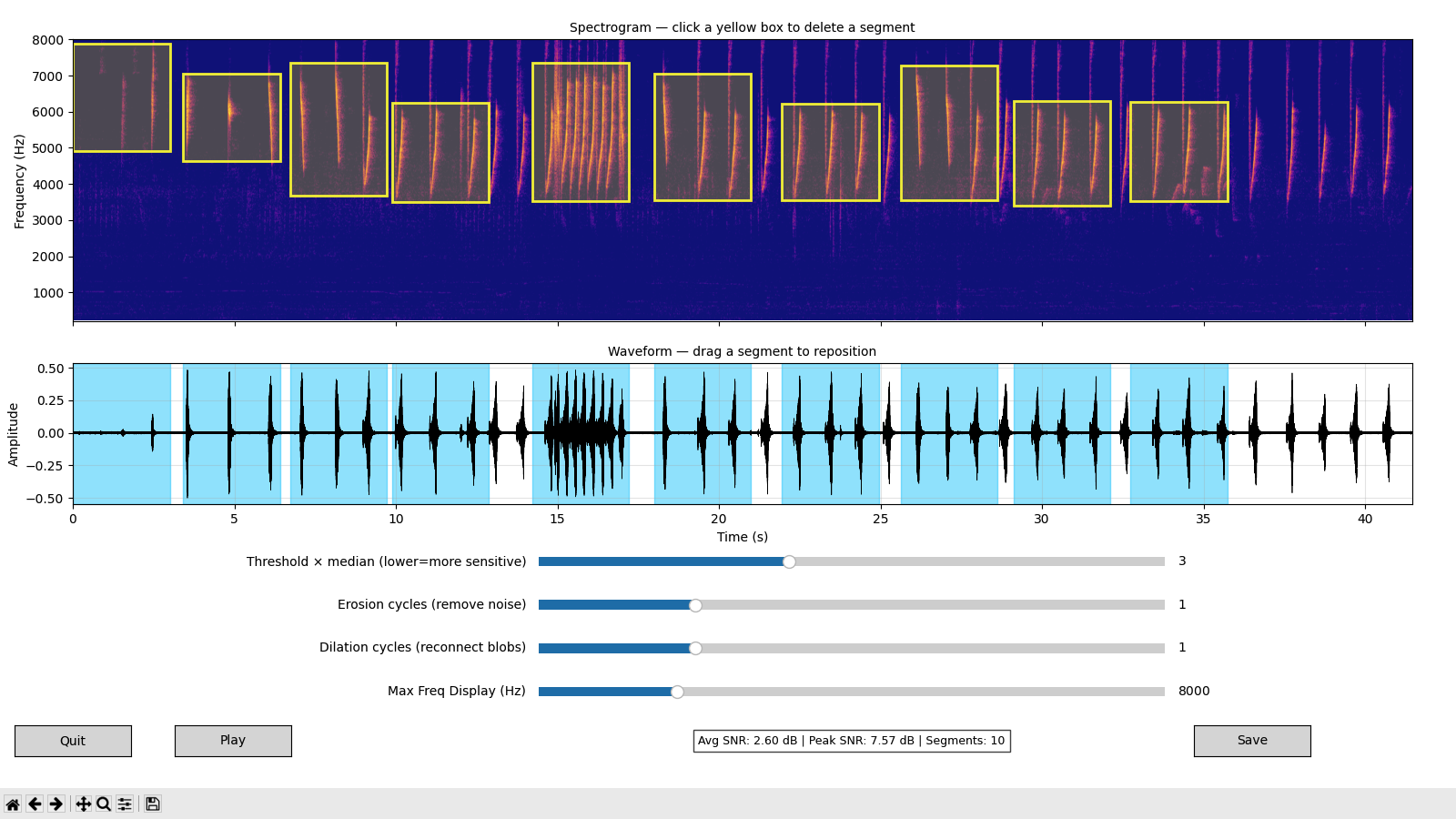}
\caption{Interactive segmentation GUI (\texttt{Stage4\_annotate\_segments.py}).
	The upper panel displays the log-mel spectrogram of the source FLAC recording
	with automatically proposed candidate segments (blob detection). The middle panel
	shows curator-approved segments as coloured overlays, each fixed at exactly three
	seconds and repositionable via drag-and-drop. The lower panel provides
	blob-detection tuning controls and audio playback functions.}
	\label{fig:annotator}
\end{figure}

To efficiently isolate target bird vocalisations, we developed a custom Python GUI (\texttt{Stage4\_annotate\_segments.py}) inspired by the spectrogram-based detection approach of Sprengel et al.\ \citep{sprengel2016audio}. As illustrated in Figure~\ref{fig:annotator}, the tool displays both the log-mel spectrogram and waveform of a source FLAC recording, while automatically proposing candidate segment boundaries using blob detection. Rather than relying solely on automated detection, the curator can interactively refine these proposals through simple drag-and-drop adjustments, remove false positives, and use the integrated playback controls to audit recordings in real time.

The GUI automatically enforces our fixed three-second segment duration, a constraint that would be incredibly tedious and time-consuming to manage manually in a standard audio editor. Once verified, the interface exports these selections as Audacity-compatible label files. Acoustically clear recordings were accepted immediately, whereas borderline cases, such as clips affected by substantial background noise or uncertain species identity, underwent a secondary review in Audacity using the exported label tracks before final inclusion in the corpus.

To keep individual, long-form recordings from dominating the dataset, we capped our extraction at a maximum of ten segments per source file. Within that limit, we actively selected for vocal diversity, prioritising distinct call types, song phrases, and alarm calls over repetitive notes to ensure models learn a robust species repertoire. This three-second duration strikes an ideal balance; it aligns with prominent bioacoustic benchmarks \citep{kahl2021birdnet} and provides plenty of acoustic real estate to capture complete, unbroken vocalisations for all twelve target species.

\begin{figure}[!htbp]
\centering
\includegraphics[width=\textwidth]{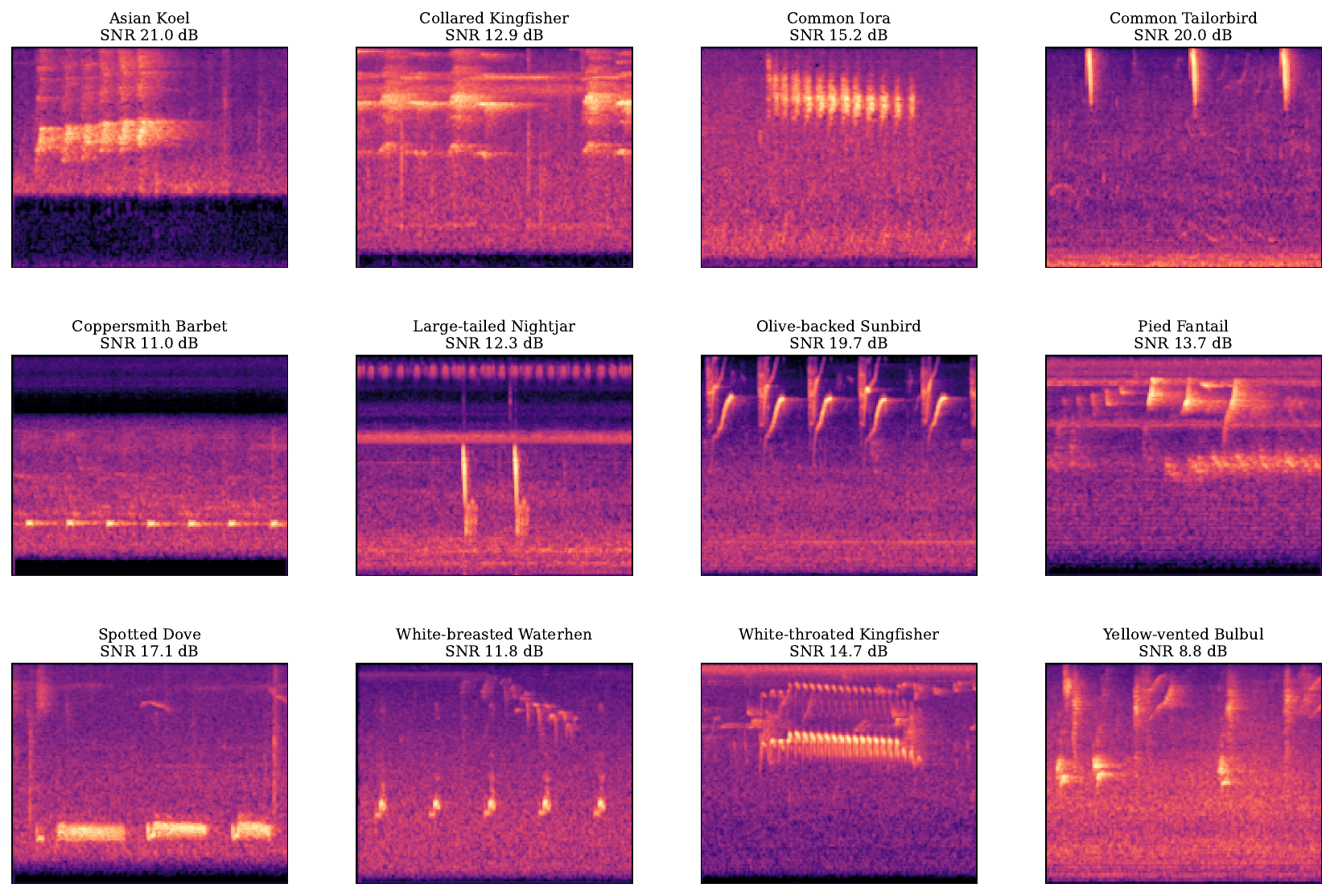}
\caption{Representative mel-spectrograms for all twelve species in MyGardenBird. Each panel shows the clip whose SNR is closest to the per-species median, providing a typical rather than best-case example. Distinctive temporal and spectral structures are visible across species, including the ascending glissando of the Asian Koel, the rapid trill of the Common Iora, and the churring nocturnal pattern of the Large-tailed Nightjar.}
\label{fig:spectrograms}
\end{figure}

We extracted these clips cleanly without any synthetic augmentation or filtering, deliberately preserving the natural acoustic environment of each environment. Out of the final corpus, our native 44.1\,kHz subset contains 6,950 clips (dropping files with lower original sampling rates), while the downsampled 16\,kHz master set retains the full 7,200 clips. To ensure complete lineage traceability back to Xeno-canto, we embedded both the original XC archive identifier and the exact clip onset time (in milliseconds) right into each WAV filename. Representative mel-spectrograms of the finalized clips are illustrated in Figure~\ref{fig:spectrograms}.

\subsection{Quality Control}

Each clip was manually reviewed and excluded if it contained overlapping calls, excessive background noise, or uncertain species identity. Rejected clips were replaced by iteratively selecting alternatives from lower-rated recordings. Clips showing digital clipping (peak amplitude $>0.99$) or low recording gain were retained as long as the target vocalisation could still be identified reliably. These recording characteristics are documented in \texttt{clips.csv} through the fields \texttt{is\_clipped}, \texttt{rms\_db}, and \texttt{peak\_amplitude}, so researchers can apply their own filtering criteria if needed.

\subsection{BirdNET Validation}

All curated clips are evaluated in zero-shot mode against BirdNET\,v2.4~\citep{kahl2021birdnet}, a large pretrained classifier covering 6{,}521 species, to provide an independent, training-free check of label integrity and clip quality before any train--test partitioning is applied. BirdNET is queried via its internal inference pipeline, bypassing the location-based species-list filter; raw sigmoid scores are restricted to the 12 target species and the top-scoring class is taken as the prediction. This step operates on all 7{,}200 clips. Full results are reported in Section~\ref{sec:validation}.

\subsection{Dataset Splitting}\label{sec:splitting}

Because multiple clips were often extracted from the same \textit{Xeno-canto} recording, a simple random split at the clip level would have scattered acoustically similar segments across the training, validation, and test sets. This would introduce a serious risk of data leakage, with models potentially learning to recognise recording‑specific background noise or microphone artifacts rather than genuine species‑level features \citep{stowell2022computational}. To avoid this bias, we required that all clips originating from a given source file remain within the same partition. We framed this grouping task as a mixed‑integer programming problem \citep{conforti2014integer}, which allowed us to minimise deviations from the target class balance while enforcing strict source‑level constraints.

Before adopting this optimisation approach, we experimented with heuristic search methods such as simulated annealing and genetic algorithms. These metaheuristics, however, proved slower to run and could not guarantee an optimal solution. By contrast, the integer programming formulation yielded a globally optimal split in under one second using the open‑source CBC solver \citep{forrest2024cbc}.

The final configuration achieves a clean 80:10:10 train/validation/test split, corresponding to 5,760 clips for training, 720 for validation, and 720 for testing. Each subset contains exactly 60 clips per species, ensuring perfect class balance across partitions. A detailed breakdown of this partitioning is provided in Table~\ref{tab:splits}. To support full reproducibility, both the generated split files and the solver implementation are included in the project repository.

\begin{table}
	\centering
	\footnotesize
	\caption{Class-wise split statistics for the 80:10:10 MIP partition. Each class contributes 480/60/60 clips to the train, validation, and test sets respectively. ``Sources'' indicates the number of unique Xeno-canto recordings in each split; the source Total column is consistent with Table~\ref{tab:species}.}
	\label{tab:splits}
	\begin{tabular}{@{}lcccccccc@{}}
		\toprule
		& \multicolumn{4}{c}{\textbf{Clips}} & \multicolumn{4}{c}{\textbf{Sources}} \\
		\cmidrule(lr){2-5} \cmidrule(lr){6-9}
		\textbf{Class} & \textbf{Train} & \textbf{Val} & \textbf{Test} & \textbf{Total} & \textbf{Train} & \textbf{Val} & \textbf{Test} & \textbf{Total} \\
		\midrule
		Asian Koel                & 480 & 60 & 60 & 600 & 102 & 12 & 22 & 136 \\
		Collared Kingfisher       & 480 & 60 & 60 & 600 & 91  & 9  & 18 & 118 \\
		Common Iora               & 480 & 60 & 60 & 600 & 74  & 10 & 21 & 105 \\
		Common Tailorbird         & 480 & 60 & 60 & 600 & 69  & 10 & 17 & 96  \\
		Coppersmith Barbet        & 480 & 60 & 60 & 600 & 90  & 11 & 19 & 120 \\
		Large-tailed Nightjar     & 480 & 60 & 60 & 600 & 78  & 10 & 20 & 108 \\
		Olive-backed Sunbird      & 480 & 60 & 60 & 600 & 78  & 11 & 16 & 105 \\
		Spotted Dove              & 480 & 60 & 60 & 600 & 74  & 10 & 13 & 97  \\
		White-breasted Waterhen   & 480 & 60 & 60 & 600 & 85  & 12 & 18 & 115 \\
		White-throated Kingfisher & 480 & 60 & 60 & 600 & 90  & 12 & 21 & 123 \\
		Yellow-vented Bulbul      & 480 & 60 & 60 & 600 & 81  & 13 & 24 & 118 \\
		Pied Fantail              & 480 & 60 & 60 & 600 & 98  & 13 & 29 & 140 \\
		\midrule
		\textbf{Total} & \textbf{5,760} & \textbf{720} & \textbf{720} & \textbf{7,200} & \textbf{1,010} & \textbf{133} & \textbf{238} & \textbf{1,381} \\
		\bottomrule
	\end{tabular}
\end{table}

\section{Data Records}\label{sec:records}

The \textbf{MyGardenBird} dataset is publicly available on Zenodo (\url{https://doi.org/10.5281/zenodo.20306877}) under a CC~BY-NC-SA~4.0 licence. It comprises 7,200 curated three‑second audio clips (600 per species across 12 species), derived from 1,381 unique \textit{Xeno-canto} source recordings. The choice of a NonCommercial--ShareAlike licence reflects the upstream terms of these recordings: 63.9\% are distributed under CC~BY-NC-SA, 35.3\% under CC~BY-NC-ND, and 0.7\% under either CC~BY or CC0. Because temporal segmentation produces derivative clips that
inherit the obligations of their parent recordings, CC~BY-NC-SA~4.0 was adopted as a consolidated licence to ensure compliance with all non‑commercial restrictions. Although CC~BY-NC-ND licences formally prohibit derivatives, we interpret the precise extraction of vocalisation boundaries as a non‑transformative curation step, consistent with established practice in bioacoustic research. This approach preserves attribution and commercial restrictions across the corpus.

To accommodate different analytical requirements, the dataset is released in two sampling‑rate configurations: \textbf{MyGardenBird16k} contains the full set of 7,200 clips downsampled to 16\,kHz ($\approx$553\,MB), while \textbf{MyGardenBird44k} retains the native 44.1\,kHz resolution for 6,950 clips
($\approx$2.0\,GB). In both versions, files are provided as standard 16‑bit PCM mono WAV tracks.

\subsection*{Repository structure}

\begin{verbatim}
mygardenbird16khz/
  Asian Koel/
    xc{source_id}_{onset_ms}.wav   e.g. xc1002657_2860.wav
  Collared Kingfisher/ ... Yellow-vented Bulbul/
mygardenbird44khz/
  <same 12 species, 44.1 kHz WAVs>
recordings.csv
metadata16khz/
  clips.csv
  qc_report.csv
  splits_mip_80_10_10.csv
metadata44khz/
  clips.csv
  qc_report.csv
  splits_mip_80_10_10.csv
mygardenbirdplus/
  16khz/
    Common Myna/   Zebra Dove/
  44khz/
    Common Myna/   Zebra Dove/
  metadata/
    16khz/
      recordings.csv   clips.csv
      qc_report.csv    splits_mip_80_10_10.csv
    44khz/
      recordings.csv   clips.csv
      qc_report.csv    splits_mip_80_10_10.csv
  README.md
mygardenbird_code/
  Stage1_xc_fetch_metadata.py ... Stage8_splitter_mip.py
  Stage9_train_mygardenbird_multifeature.py
  config.py   requirements.txt
\end{verbatim}

Species folders use the English common name (e.g.\ \texttt{Asian Koel/}, \texttt{Collared Kingfisher/}). The WAV filename encodes the Xeno-canto source identifier and clip start time in milliseconds from the beginning of the source FLAC file, enabling exact traceability back to the original recording. The \texttt{mygardenbirdplus/} directory contains 600 clips per species for Common Myna and Zebra Dove at both sampling rates. Both species were excluded from the core dataset because only 75 and 577 clips respectively are available from ASEAN/Indo-Malayan regional sources (see Section~2.1); the addendum supplements these with global Xeno-canto recordings to reach exactly 600 clips per species for each, and is intended for applications where strict regional provenance is not required. Its \texttt{metadata/16khz/} and \texttt{metadata/44khz/} subfolders each provide the matching \texttt{recordings.csv}, \texttt{clips.csv}, \texttt{qc\_report.csv}, and \texttt{splits\_mip\_80\_10\_10.csv} for the respective rate.

\subsection*{Metadata schema}

Metadata is distributed as three normalised CSV files whose fields are described in Tables~\ref{tab:recordings} and~\ref{tab:clips}.

\begin{itemize}
	\item \texttt{recordings.csv} (archive root; one row per Xeno-canto source recording, primary key \texttt{source\_id}) holds provenance attributes.
	\item \texttt{metadata16khz/clips.csv} and \texttt{metadata44khz/clips.csv} (one row per 3-second WAV clip, primary key \texttt{file\_id}) hold per-clip signal-quality metrics and link each clip back to its source recording via \texttt{source\_id}.
\end{itemize}

\begin{table}[h]
\centering\small
\caption{Fields in \texttt{recordings.csv} (one row per source recording; PK: \texttt{source\_id}).}
\label{tab:recordings}
\begin{tabular}{@{}lll@{}}
\toprule
\textbf{Field} & \textbf{Type} & \textbf{Description} \\
\midrule
\texttt{source\_id}          & string  & Xeno-canto recording identifier (PK) \\
\texttt{species\_common}     & string  & English common name \\
\texttt{species\_scientific} & string  & Binomial scientific name \\
\texttt{quality\_grade}      & string  & Xeno-canto recorder quality grade (A--E) \\
\texttt{cc\_license}         & string  & Creative Commons licence SPDX identifier \\
\texttt{type\_label}         & string  & Normalised vocalisation type: song / call / other \\
\texttt{latitude}            & float   & Recording latitude (WGS84; blank if unknown) \\
\texttt{longitude}           & float   & Recording longitude (WGS84; blank if unknown) \\
\texttt{country}             & string  & Country of recording (ISO name from XC) \\
\bottomrule
\end{tabular}
\end{table}

\begin{table}[h]
\centering\small
\caption{Fields in \texttt{clips.csv} (in \texttt{metadata16khz/} or \texttt{metadata44khz/}; $n=7{,}200$ at 16\,kHz, $n=6{,}950$ at 44.1\,kHz; same schema for both; PK: \texttt{file\_id}; FK: \texttt{source\_id}).}
\label{tab:clips}
\begin{tabular}{@{}lll@{}}
\toprule
\textbf{Field} & \textbf{Type} & \textbf{Description} \\
\midrule
\texttt{file\_id}            & string  & Unique clip identifier: \texttt{xc\{source\_id\}\_\{onset\_ms\}} (PK) \\
\texttt{source\_id}          & string  & Xeno-canto recording identifier (FK $\to$ \texttt{recordings.csv}) \\
\texttt{onset\_ms}         & integer & Clip onset in milliseconds from start of source FLAC \\
\texttt{sampling\_rate}      & integer & Clip sampling rate (Hz) \\
\texttt{snr\_db}             & float   & Estimated SNR (dB); percentile-based noise floor \\
\texttt{rms\_db}             & float   & Whole-clip RMS level (dBFS) \\
\texttt{peak\_amplitude}     & float   & Peak sample amplitude (0--1) \\
\texttt{is\_clipped}         & boolean & True if peak amplitude $> 0.99$ \\
\bottomrule
\end{tabular}
\end{table}

\subsection*{Split file}

The file \texttt{metadata16khz/splits\_mip\_80\_10\_10.csv} (and its counterpart in \texttt{metadata44khz/}) assigns every clip to one of three partitions (\texttt{train} / \texttt{val} / \texttt{test}) at an 80:10:10 ratio, with assignment performed at the source-recording level to prevent leakage (Section~\ref{sec:splitting}). It contains two columns: \texttt{file\_id} (foreign key referencing \texttt{metadata16khz/clips.csv}) and \texttt{split}. The released split is fixed; the CSV can be used directly without re-running the solver. Users who re-run the solver may obtain a different equally-optimal assignment depending on source ordering, as the MIP objective equals zero for all such assignments.

\subsection*{Relational structure}

\begin{figure}[!htbp]
\centering
\includegraphics[width=0.66\textwidth]{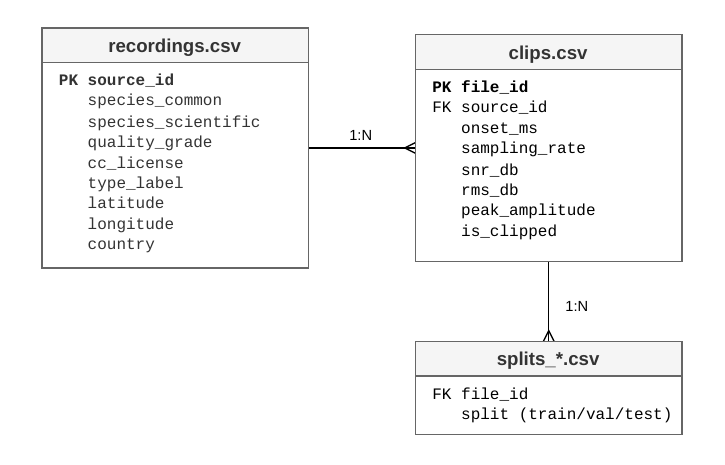}
\caption{Entity--Relationship diagram for the MyGardenBird metadata files. \texttt{recordings.csv} (primary key \texttt{source\_id}; one row per source recording) captures provenance attributes including species identity, geographic coordinates, quality grade, and Creative Commons licence (\texttt{cc\_license} as an SPDX identifier), and links via a one-to-many relationship to \texttt{metadata16khz/clips.csv} (or \texttt{metadata44khz/clips.csv}; primary key \texttt{file\_id} = \texttt{xc\{source\_id\}\_\{onset\_ms\}}, foreign key \texttt{source\_id}; one row per 3-second clip). Each \texttt{clips.csv} in turn links to the corresponding split file \texttt{metadata16khz/splits\_mip\_80\_10\_10.csv} (or \texttt{metadata44khz/}; foreign key \texttt{file\_id}; one row per clip), whose \texttt{split} field takes values \texttt{train}, \texttt{val}, or \texttt{test}. This schema provides complete traceability from each audio clip back to its Xeno-canto source recording and its assigned dataset partition.}
\label{fig:er}
\end{figure}

The three distributed CSV files follow the relational schema illustrated in Figure~\ref{fig:er}. \texttt{recordings.csv} (archive root) holds one row per Xeno-canto source recording (primary key: \texttt{source\_id}) and captures provenance attributes including species identity, geographic coordinates, quality grade, Creative Commons licence (\texttt{cc\_license} as an SPDX identifier), and country of origin. Each source recording maps to one or more clips in \texttt{metadata16khz/clips.csv} or \texttt{metadata44khz/clips.csv} (primary key: \texttt{file\_id}, foreign key: \texttt{source\_id}), which stores per-clip signal-quality metrics (SNR, RMS, peak amplitude) and the clip onset time. Split membership is recorded in \texttt{metadata16khz/splits\_mip\_80\_10\_10.csv} (or \texttt{metadata44khz/}) via \texttt{file\_id} and a \texttt{split} field taking values \texttt{train}, \texttt{val}, or \texttt{test}. This three-table design keeps provenance, signal-quality, and partition information separately queryable while preserving full traceability from each clip back to its source recording.

\section{Technical Validation}\label{sec:validation}

Dataset integrity was evaluated at three levels: (i) objective signal-quality metrics derived from programmatic inspection of all 7,200 clips, (ii) training-free evaluation against BirdNET\,v2.4 as a same-source label-consistency check across the full dataset, and (iii) classification consistency across heterogeneous CNN architectures on held-out test clips, used as a proxy for class separability rather than as a benchmarking study.

\subsection{Signal quality}

\begin{figure}[!htbp]
\centering
\includegraphics[width=\textwidth]{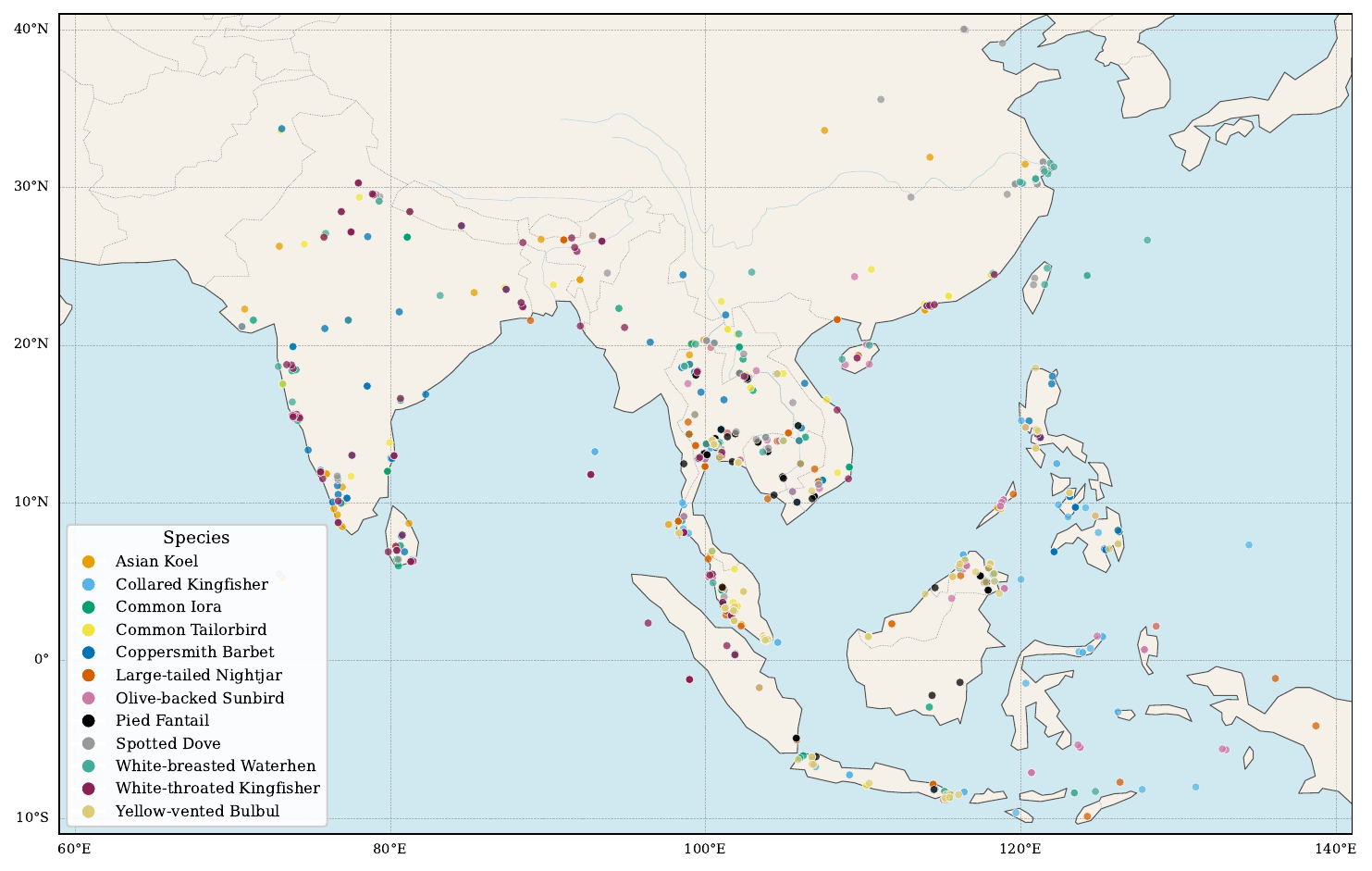}
\caption{Geographic distribution of the Xeno-canto source recordings used in MyGardenBird (sources with available geographic coordinates shown). Each point represents one source recording; colours denote species. Recordings span Peninsular Malaysia, neighbouring Southeast Asian countries, and the broader South and East Asian range of each species. For species with limited regional coverage, recordings from South Asian and East Asian countries were incorporated to reach the 600-clip target; all twelve species are native residents of Peninsular Malaysia and the broader Indo-Malayan biogeographic region.}
\label{fig:map}
\end{figure}

\begin{figure}[!htbp]
\centering
\includegraphics[width=\textwidth]{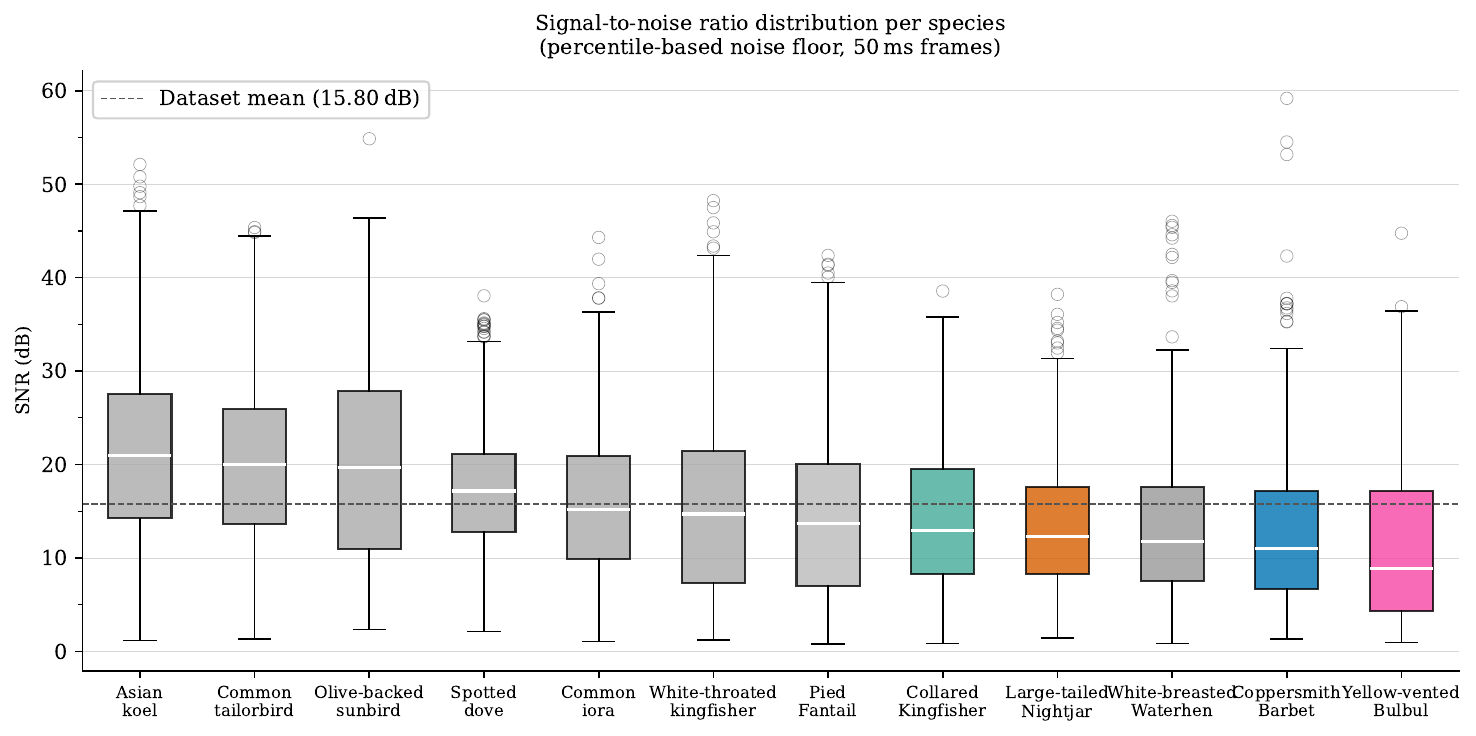}
\caption{Per-species signal-to-noise ratio distribution. SNR was estimated using a percentile-based noise-floor method (10th-percentile of 50\,ms frame-level RMS values). Box plots show the interquartile range; whiskers extend to 1.5$\times$ IQR; circles denote outliers. The dashed line marks the dataset-wide mean of 15.80\,dB. Most species exhibit median SNR above 10\,dB; Yellow-vented Bulbul has the lowest median (8.90\,dB), reflecting its prevalence in noisier urban environments.}
\label{fig:snr}
\end{figure}

SNR was estimated using a percentile-based noise-floor method (10th-percentile of 50\,ms frame-level RMS values). This estimator approximates the noise floor as the lower tail of the short-term energy distribution, under the assumption that vocalisation-active frames are a minority of the three-second clip; it is robust to brief high-energy noise events (e.g.\ wind gusts, distant traffic) common in garden recordings, as these affect few frames without substantially shifting the lower percentile. Across all 7,200 clips, SNR ranged from 0.83 to 59.18\,dB (mean 15.80\,dB, s.d.\ 9.10\,dB); per-species means ranged from approximately 11 to 21\,dB (Figure~\ref{fig:snr}). Fewer than 2\% of clips exceeded the digital clipping threshold (peak amplitude $>0.99$). All clips are exactly 3.000\,s at 16\,kHz; no corrupt or unreadable files were found. Geographic coordinates are available for 1,325 of 1,381 source recordings (95.9\%), spanning 20 countries across the Indo-Malayan region (Figure~\ref{fig:map}).

\subsection{Structural integrity}

Class balance was verified programmatically: all twelve species contribute exactly 600 clips, confirmed by counting rows in \texttt{metadata16khz/clips.csv}. No duplicate \texttt{file\_id} values or duplicate (\texttt{source\_id}, \texttt{clip\_index}) pairs were found. \texttt{metadata16khz/splits\_mip\_80\_10\_10.csv} was cross-checked to confirm zero overlap between train, validation, and test sets at the source-recording level; no \texttt{source\_id} appears in more than one partition.

\subsection{Annotation protocol}

All clips were annotated by a single domain-expert annotator using a consistent two-stage protocol: (i) automated candidate segment detection using the spectrogram blob method of Sprengel et al.\ \citep{sprengel2016audio}, followed by (ii) manual review and correction of every segment in Audacity using simultaneous auditory and spectrogram inspection. Segments were accepted only when the target vocalisation was clearly audible, species identity was unambiguous, and no severe overlapping calls or background interference obscured the signal. Annotations were not carried over across recording sessions; each recording was reviewed in full before any segment was accepted.

Because all annotations were produced by a single annotator, no inter-rater reliability measure (e.g.\ Cohen's $\kappa$) is available. The internal consistency of the annotation process is instead supported indirectly by the BirdNET label-consistency results (Section~\ref{sec:validation}) and by the low seed-to-seed variance observed in the CNN experiments (s.d.\ $<1.0\%$ across all architectures), both of which are sensitive to systematic labelling errors. Single-annotator curation is a recognised limitation of this dataset; future releases should incorporate independent second-annotator verification on a random subsample.

\subsection{Training-free external validation with BirdNET}

The two validation roles served by this section are distinct: BirdNET evaluates \emph{label consistency} at the dataset level, whereas the CNN experiments in Section~\ref{sec:cnn} measure \emph{class separability} on held-out clips. Because BirdNET\,v2.4~\citep{kahl2021birdnet} is a fixed pretrained classifier covering 6{,}521 species and is not fitted on MyGardenBird in any way, applying it to all 7{,}200 clips introduces no data leakage and provides a statistically stronger estimate of annotation consistency than restricting evaluation to the test partition alone. BirdNET is therefore evaluated on the complete 16\,kHz dataset; CNN models use the 80:10:10 split as described in Section~\ref{sec:cnn}.

One limitation of this evaluation is that BirdNET\,v2.4 was also trained using recordings from Xeno-canto~\citep{kahl2021birdnet}, which is the same source archive used for MyGardenBird. As a result, strong agreement between BirdNET predictions and MyGardenBird labels mainly indicates consistency with the broader Xeno-canto corpus, showing that the curated clips are representative and not systematically mislabelled within that source. It should not be interpreted as fully independent validation from a model trained on a separate dataset. Even so, we consider this a useful quality-control step, while recognising that validation with a model trained on non--Xeno-canto recordings would provide stronger evidence of label accuracy.

BirdNET was queried via its internal inference pipeline directly, bypassing the high-level analysis API (which applies a location-based species-list filter that restricts candidate species to those expected at a given GPS location; without a location argument, the filter suppresses predictions for all species not on a globally ubiquitous default list, making it unsuitable for pre-segmented clips from a fixed species set). For each 3-second clip, raw sigmoid scores across all 6{,}521 classes were obtained; scores were restricted to the 12 MyGardenBird species and the top-scoring class was taken as the prediction.

BirdNET achieves \textbf{97.94\%} accuracy at 16\,kHz (95\,\% CI: 97.59\%--98.25\%; $n$\,=\,7{,}200; macro AUC\,=\,0.9913) and \textbf{98.06\%} at 44.1\,kHz (95\,\% CI: 97.71\%--98.36\%; $n$\,=\,6{,}950; macro AUC\,=\,0.9922) with no fine-tuning. Per-species results for both subsets are shown in Table~\ref{tab:birdnet}. No species falls below F1\,=\,0.97 at either sampling rate. These results confirm that: (i) clip labels are consistent with BirdNET's independently derived species taxonomy, (ii) vocalisations are sufficiently clean and species-typical to be identifiable by a general-purpose model, and (iii) the dataset contains no systematic mislabelling or species confusion that would suppress recognition by an external system.

The high accuracy is in line with BirdNET's training data, which includes all twelve target species and a substantial number of recordings from Asia. The lower PR AUC values reported in global evaluations on field recordings \citep{funosas2026global} are better understood as a consequence of working with uncurated, species-rich soundscapes rather than a limitation of the model itself. When applied to manually reviewed, precisely segmented clips of acoustically distinct species, BirdNET performs close to its expected upper bound.

Alongside the shared provenance caveat mentioned earlier, these findings suggest that the curated clips are consistent with the broader Xeno-canto corpus. They appear species typical, cleanly segmented, and free from systematic labelling issues, rather than serving as fully independent confirmation of label correctness.

\begin{table}
\centering
\caption{BirdNET\,v2.4 training-free external validation on all clips in each subset (16\,kHz: $n$\,=\,7{,}200, 600 per species; 44.1\,kHz: $n$\,=\,6{,}950, 570--599 per species). No fine-tuning; prediction is the top-scoring class among the 12 target species. Overall accuracy: 97.94\% at 16\,kHz (95\,\% CI: 97.59\%--98.25\%); 98.06\% at 44.1\,kHz (95\,\% CI: 97.71\%--98.36\%).}
\label{tab:birdnet}
\smallskip
\small
\begin{tabular}{@{}lcccccc@{}}
\toprule
& \multicolumn{3}{c}{\textbf{16\,kHz}} & \multicolumn{3}{c}{\textbf{44.1\,kHz}} \\
\cmidrule(lr){2-4}\cmidrule(lr){5-7}
\textbf{Species} & \textbf{F1} & \textbf{AUC} & \textbf{Rec} & \textbf{F1} & \textbf{AUC} & \textbf{Rec} \\
\midrule
Asian Koel                & 0.983 & 0.981 & 0.982 & 0.983 & 0.985 & 0.986 \\
Collared Kingfisher       & 0.985 & 0.999 & 0.983 & 0.987 & 0.999 & 0.983 \\
Common Iora               & 0.973 & 0.974 & 0.983 & 0.971 & 0.978 & 0.958 \\
Common Tailorbird         & 0.972 & 0.988 & 0.965 & 0.972 & 0.977 & 0.958 \\
Coppersmith Barbet        & 0.985 & 0.996 & 0.985 & 0.983 & 0.997 & 0.977 \\
Large-tailed Nightjar     & 0.989 & 1.000 & 1.000 & 0.991 & 1.000 & 0.998 \\
Olive-backed Sunbird      & 0.970 & 0.990 & 0.982 & 0.971 & 0.996 & 0.993 \\
Pied Fantail              & 0.975 & 0.992 & 0.957 & 0.971 & 0.990 & 0.976 \\
Spotted Dove              & 0.980 & 0.991 & 0.978 & 0.985 & 0.993 & 0.983 \\
White-breasted Waterhen   & 0.982 & 0.990 & 0.972 & 0.986 & 0.996 & 0.983 \\
White-throated Kingfisher & 0.985 & 0.998 & 0.985 & 0.987 & 0.998 & 0.990 \\
Yellow-vented Bulbul      & 0.974 & 0.998 & 0.982 & 0.980 & 0.998 & 0.983 \\
\midrule
\textbf{Macro avg}        & \textbf{0.979} & \textbf{0.991} & \textbf{0.979} & \textbf{0.981} & \textbf{0.992} & \textbf{0.981} \\
\bottomrule
\end{tabular}
\end{table}

\subsection{Classification consistency}\label{sec:cnn}

Three CNN architectures were fine-tuned on 224$\times$224 log-Mel spectrograms (N\_FFT\,=\,2048; 224 mel bins; hop\,=\,$\lfloor N_\text{samples}/224 \rfloor$ where $N_\text{samples} = \text{sampling rate} \times \text{clip duration}$, e.g.\ $16{,}000 \times 3 = 48{,}000$ at 16\,kHz, giving 214\,samples at 16\,kHz and 590\,samples at 44.1\,kHz) using AdamW ($lr$=$10^{-3}$, weight decay=$10^{-5}$), batch size 32, early stopping (patience 10), and Mixup \citep{zhang2018mixup} ($\alpha$=0.2). Each model was initialised with ImageNet weights and trained with three random seeds (42, 100, 786) on the 80:10:10 source-separated split. The architectures span a range of computational budgets: MobileNetV3-Small \citep{howard2019searching} (2.9\,M parameters, 56\,MFLOPs), EfficientNet-B0 \citep{tan2019efficientnet} (5.3\,M, 390\,MFLOPs), and ResNet-50 \citep{he2016deep} (25.6\,M, 4.1\,GFLOPs). The hop length is set so that a 3-second clip produces exactly 224--225 time frames (with librosa \texttt{center=True} padding); the resulting feature map is cropped to 224$\times$224 without interpolation and replicated across three channels to match ImageNet input conventions. Consistent accuracy across this heterogeneous set is interpretable as evidence that class boundaries are acoustically well-defined; architecture benchmarking is not the primary objective.

\subsection{Results and interpretation}

Table~\ref{tab:accuracies} summarises test accuracy for all three CNN architectures and the BirdNET\,v2.4 external reference, for both 16\,kHz and 44.1\,kHz subsets.

\begin{table}[bh]
\centering
\caption{Classification accuracy (\%) for CNN models and BirdNET\,v2.4. CNN results: mean $\pm$ s.d. across three seeds (42, 100, 786), Mel-spectrogram features, ImageNet initialisation, Mixup ($\alpha$=0.2), 80:10:10 test split (720/695 clips at 16/44.1\,kHz). BirdNET\,v2.4 (no fine-tuning): evaluated on all clips in each subset (7{,}200 at 16\,kHz; 6{,}950 at 44.1\,kHz) --- dataset-level label validation, no leakage risk.}
\label{tab:accuracies}
\smallskip
\small
\begin{tabular}{@{}lcc@{}}
\toprule
\textbf{Model} & \textbf{16\,kHz} & \textbf{44.1\,kHz} \\
\midrule
MobileNetV3-Small & $92.41 \pm 0.64$ & $90.70 \pm 0.30$ \\
EfficientNet-B0   & $\mathbf{96.39 \pm 0.69}$ & $\mathbf{94.24 \pm 0.77}$ \\
ResNet-50         & $94.63 \pm 0.07$ & $93.09 \pm 0.62$ \\
\midrule
BirdNET\,v2.4 (no fine-tuning) & $97.94$ & $98.06$ \\
\bottomrule
\end{tabular}
\end{table}

\paragraph{Performance and Augmentation Trends}

Table~\ref{tab:aug} breaks down the 16\,kHz test accuracies for all three architectures across our different augmentation setups. Incorporating SpecAugment \citep{park2019specaugment} provided a clear boost for MobileNetV3-Small (+2.87\,pp) and ResNet-50 (+1.01\,pp), though it flatlined for EfficientNet-B0. This trend suggests that aggressive frequency and time masking offers the greatest returns for models with lower baseline capacity. Meanwhile, Mixup ($\alpha$=0.2) proved to be our most powerful and consistent regularizer across the board. It comfortably outperformed the unaugmented baseline across all three networks---yielding improvements of 3.01\,pp for MobileNetV3-Small, 1.48\,pp for EfficientNet-B0, and 1.57\,pp for ResNet-50. Given its strong, uniform performance, we adopted Mixup as our default augmentation strategy for the core benchmarks reported in Table~\ref{tab:accuracies}.

\begin{table}[h]
\centering
\caption{16\,kHz test accuracy (\%) for three augmentation strategies. Mean\,$\pm$\,s.d.\ across three seeds (42, 100, 786); 80:10:10 source-separated split (720 test clips).}
\label{tab:aug}
\smallskip
\small
\begin{tabular}{@{}lccc@{}}
\toprule
\textbf{Model} & \textbf{No augmentation} & \textbf{SpecAugment} & \textbf{Mixup ($\alpha$=0.2)} \\
\midrule
MobileNetV3-Small & $89.40 \pm 1.35$ & $92.27 \pm 0.17$ & $\mathbf{92.41 \pm 0.64}$ \\
EfficientNet-B0   & $94.91 \pm 0.85$ & $94.91 \pm 0.85^{\dagger}$ & $\mathbf{96.39 \pm 0.69}$ \\
ResNet-50         & $93.06 \pm 0.69$ & $94.07 \pm 0.47$ & $\mathbf{94.63 \pm 0.07}$ \\
\bottomrule
\end{tabular}
\par\smallskip
{\footnotesize $^{\dagger}$Identical to No-augmentation by numerical coincidence; per-seed values differ (No-aug: 94.31, 94.31, 96.11\%; SpecAugment: 94.86, 93.89, 95.97\%).}
\end{table}

\begin{figure}[!htbp]
\centering
\includegraphics[width=0.6\textwidth]{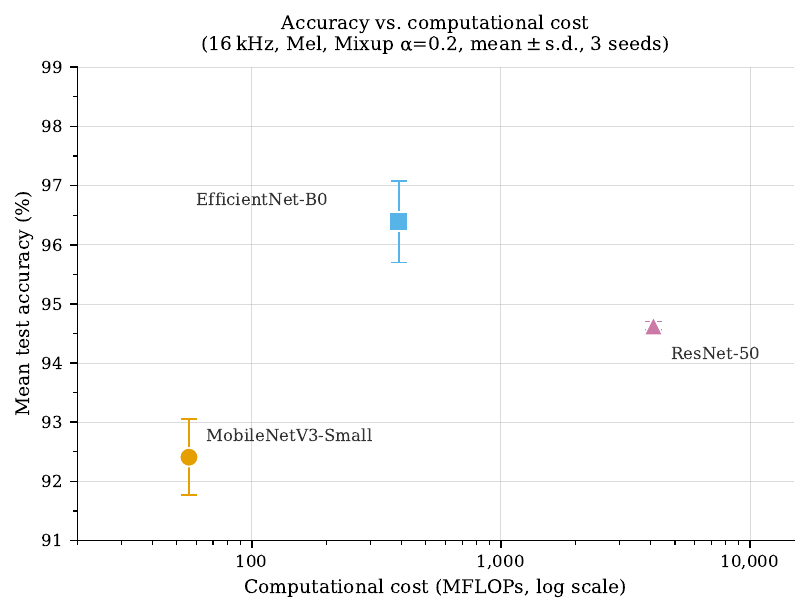}
\caption{Mean test accuracy versus computational cost (MFLOPs, log scale) for the three CNN architectures trained on 16\,kHz Mel-spectrograms (Mixup $\alpha$=0.2, ImageNet initialisation; mean\,$\pm$\,s.d.\ across three random seeds). MobileNetV3-Small (56\,MFLOPs, $92.41\pm0.64$\%) is the most compute-efficient option; EfficientNet-B0 (390\,MFLOPs, $96.39\pm0.69$\%) achieves the highest accuracy; ResNet-50 (4{,}100\,MFLOPs, $94.63\pm0.07$\%) occupies an intermediate position and exhibits the smallest seed-to-seed variance. The narrow error bars across a 73$\times$ compute range ($56$--$4{,}100$\,MFLOPs) indicate stable training dynamics; consistent 92--96\% accuracy across architectures of such varied capacity is consistent with acoustically well-separated class boundaries in the dataset.}
\label{fig:accuracy-vs-compute}
\end{figure}

When trained at 16\,kHz with Mixup augmentation, EfficientNet-B0 achieved the highest mean accuracy ($96.39 \pm 0.69\%$), followed by ResNet-50 ($94.63 \pm 0.07\%$) and MobileNetV3-Small ($92.41 \pm 0.64\%$). Figure~\ref{fig:accuracy-vs-compute} illustrates the trade-off between classification performance and computational complexity across the evaluated architectures. Notably, all models exhibited low variability across random seed initialisations, with standard deviations below $1.0\%$. This consistency indicates stable optimisation dynamics and suggests that the dataset supports reliable learning of discriminative species-level acoustic features. These baseline results are presented primarily as an assessment of dataset quality and experimental reproducibility, rather than as a comparative evaluation intended to identify a superior model architecture.

A similar pattern was observed for the 44.1\,kHz subset, where Mixup consistently outperformed both SpecAugment and the unaugmented baseline across all evaluated architectures. However, all three models exhibited a modest reduction in accuracy relative to their corresponding 16\,kHz results under Mixup, with decreases of 2.15 percentage points for EfficientNet-B0, 1.71 for MobileNetV3-Small, and 1.54 for ResNet-50.

Direct comparison between the two subsets should be interpreted with caution because they differ in several respects. First, the test partitions are not identical in size: the MIP-optimised 80:10:10 split produced 720 test clips for the 16\,kHz dataset and 695 test clips for the 44.1\,kHz dataset, reflecting the smaller overall pool of recordings available at the higher sampling rate (6,950 versus 7,200 clips). Second, the 44.1\,kHz subset contains only recordings natively acquired at that sampling rate, thereby excluding lower-bandwidth recordings. Consequently, the 44.1\,kHz subset represents a more selective collection of recordings rather than a lower-quality one.

One possible explanation for the observed performance difference relates to spectrogram representation. Because all models used a fixed input resolution of 224$\times$224 pixels, the higher sampling rate required a proportionally larger hop length during spectrogram generation. This adjustment reduces temporal resolution and may limit the representation of short-duration acoustic events that contribute to species discrimination. Nevertheless, because the 44.1\,kHz subset draws from only 1,327 of the 1,381 source recordings rather than the complete pool (the remaining 54 were natively below 44.1\,kHz and appear only in the 16\,kHz set), the test partitions are not matched, and the present results do not constitute a controlled ablation of sampling rate. Further experimentation using the same recordings processed at both rates would be required to isolate the effect of sampling rate from that of recording selection.

Importantly, the magnitude of the reduction remained small, not exceeding 2.2 percentage points for any architecture, and the effect was observed consistently across all three models. This consistency suggests that the phenomenon is more likely attributable to characteristics of the dataset and preprocessing pipeline than to architecture-specific behaviour.

\paragraph{Species-Level Analysis}

\begin{figure}[!htbp]
	\centering
	\includegraphics[width=\textwidth]{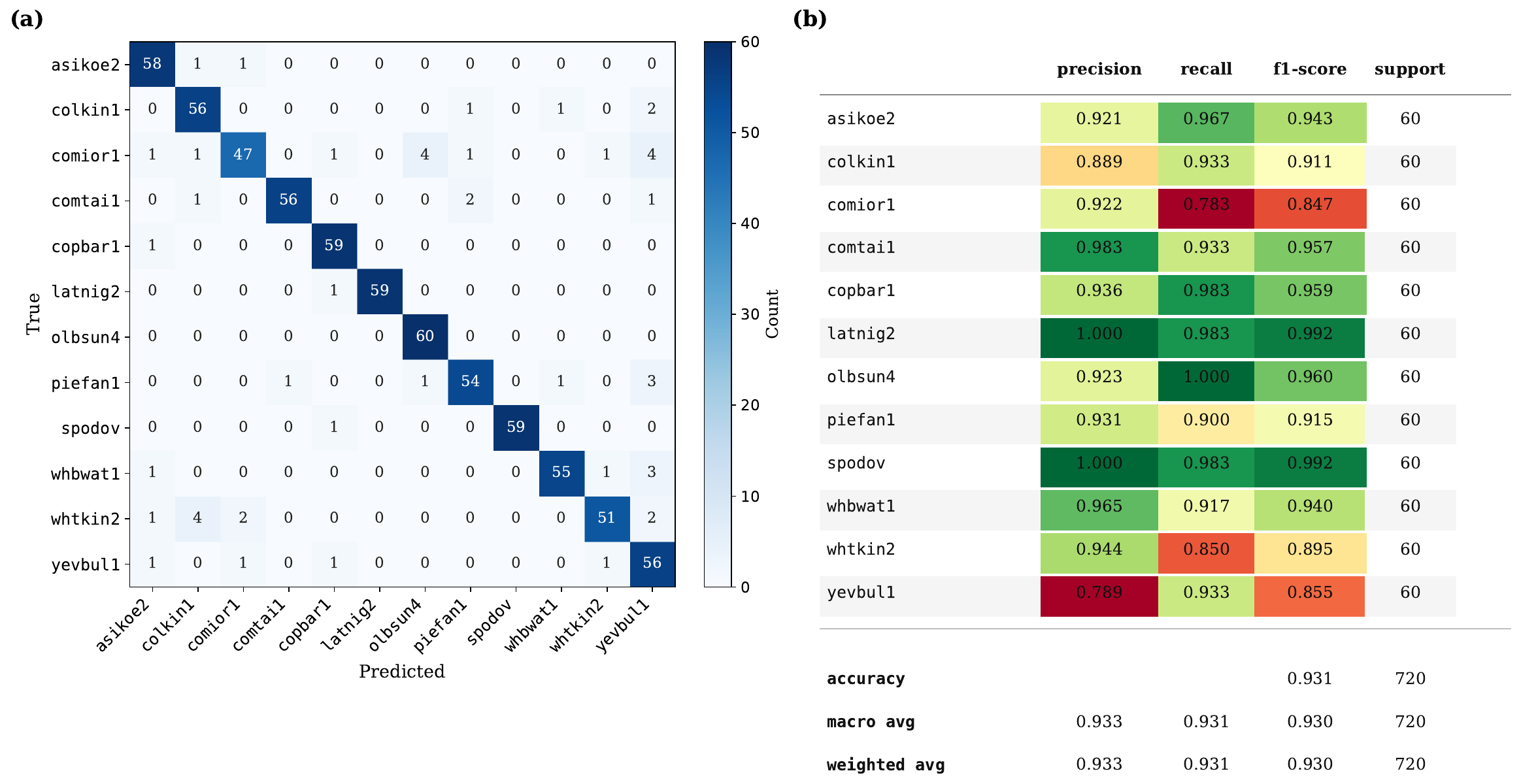}
	\caption{Per-class performance of MobileNetV3-Small on the 16\,kHz test set (Mixup $\alpha=0.2$, seed = 42; 60 clips per class). (a) Confusion matrix; (b) class-wise precision, recall, and F1-score. Species are identified using eBird taxon codes: asikoe2 = Asian Koel, colkin1 = Collared Kingfisher, comior1 = Common Iora, comtai1 = Common Tailorbird, copbar1 = Coppersmith Barbet, latnig2 = Large-tailed Nightjar, olbsun4 = Olive-backed Sunbird, piefan1 = Pied Fantail, spodov = Spotted Dove, whbwat1 = White-breasted Waterhen, whtkin2 = White-throated Kingfisher, yevbul1 = Yellow-vented Bulbul.}
	\label{fig:confusion}
\end{figure}

Figure~\ref{fig:confusion} presents the per-class performance of MobileNetV3-Small (trained with Mixup $\alpha=0.2$, seed = 42) on the 16\,kHz test set. This architecture was selected for detailed species-level analysis due to its comparatively lower representational capacity, which provides a more conservative indication of class separability by exposing ambiguities that may be attenuated in higher-capacity models.

Overall, the confusion matrix exhibits a sparse distribution of off-diagonal errors, indicating generally well-separated class boundaries. The most prominent error patterns are biologically interpretable. The first involves the Common Iora, which accounts for 13 misclassified clips distributed primarily across acoustically similar species such as the Olive-backed Sunbird and Yellow-vented Bulbul. The second involves the White-throated Kingfisher, with 9 misclassifications occurring predominantly within its congeners, particularly the Collared Kingfisher. At the other extreme, the Olive-backed Sunbird achieved perfect classification performance (recall = 1.000). Across all classes, the Common Iora yielded the lowest F1-score (0.847), whereas the Large-tailed Nightjar and Spotted Dove achieved the highest F1-scores (0.992 each).

The relatively lower performance of the Common Iora is consistent with its highly diverse vocal repertoire \citep{rasmussen2012birds,divyapriya2019spectral}. Males of this species produce multiple distinct phrase types, and given the source-level splitting strategy, certain vocal variants present in the test set may not have been represented in the training set. This mismatch likely contributes to reduced generalization performance for this class under a lightweight architecture.

The White-breasted Waterhen exhibits a contrasting pattern, with near-perfect recall at 16\,kHz but reduced performance at 44.1\,kHz. One plausible explanation is that the higher sampling rate preserves high-frequency insect stridulation ($>$8\,kHz) common in the species' wetland habitats, which are attenuated in the 16\,kHz representation. The resulting increase in background spectral complexity may introduce confounding features that are otherwise suppressed in the band-limited setting.

Figure~\ref{fig:accuracy-vs-compute} places these results in a computational context. Despite spanning a 73$\times$ range in inference cost, all three architectures achieve 92--96\% accuracy with seed-to-seed variance below 1\,pp --- a spread that is small both in absolute terms and relative to the compute gap between models. That a lightweight 56\,MFLOP mobile network and a 4.1\,GFLOP deep residual network land within four percentage points of each other, with similarly tight confidence intervals, is the central finding of this section: the acoustic class structure encoded in MyGardenBird is robust enough to be consistently and reliably learned across architectures of substantially different capacity.

\section*{Usage Notes}

Both subsets are organised into species-specific subfolders. Filenames contain the Xeno-canto recording ID and the clip start time in milliseconds (e.g.\ \texttt{xc1002657\_2860.wav}). Species labels and geographic metadata are provided in \texttt{recordings.csv} (archive root), while per-clip SNR and quality measurements are listed in \texttt{metadata16khz/clips.csv} or \texttt{metadata44khz/clips.csv}. Train/validation/test split assignments are stored in \texttt{metadata16khz/splits\_mip\_80\_10\_10.csv} (or \texttt{metadata44khz/}); see Tables~\ref{tab:recordings}--\ref{tab:clips}.

The 16\,kHz subset is intended primarily for edge-AI applications. A MobileNetV3-Small model quantised to 8-bit occupies roughly 1.5\,MB under TensorFlow Lite integer quantisation, making it practical for smartphones and resource-constrained devices such as the ESP32-S3. Larger models, including EfficientNet-B0 and ResNet-50, are better suited to single-board computers such as the Raspberry Pi\,4/5 or Jetson Nano/Orin. The 44.1\,kHz subset is designed for higher-resolution spectro-temporal analysis and transfer-learning experiments. Researchers can also apply additional quality filtering using fields such as \texttt{is\_clipped}, \texttt{rms\_db}, and \texttt{peak\_amplitude} from \texttt{metadata16khz/clips.csv} or \texttt{metadata44khz/clips.csv}.

\section*{Code Availability}

All curation, preprocessing, and model training code (Python~3.10, TensorFlow~2.15, Librosa) is available at \url{https://github.com/mun3im/mygardenbird} with a \texttt{requirements.txt} for full reproducibility, and is persistently archived within the Zenodo repository at \url{https://doi.org/10.5281/zenodo.20306877}.

\section*{Acknowledgements}

This work was supported by Universiti Malaya and benefited from the contributions of the open Xeno-canto community. The authors thank the contributors of the Creative Commons--licensed recordings that made this dataset possible.

\section*{Author Contributions}

M.M.A.Z. conceived the dataset, curated and processed the recordings, conducted the model development and experiments, and drafted the initial manuscript. M.Y.I.I. and N.I. provided guidance on the study design, methodology, and manuscript preparation. All authors reviewed, revised, and approved the final version of the manuscript.

\section*{Competing Interests}

The authors declare no competing interests.

\bibliography{sn-bibliography}

@book{conforti2014integer,
	title        = {Integer Programming},
	author       = {Conforti, Michele and Cornu{\'e}jols, G{\'e}rard and Zambelli, Giacomo},
	year         = {2014},
	publisher    = {Springer},
	address      = {Cham},
	doi          = {10.1007/978-3-319-11008-0}
}

@article{divyapriya2019spectral,
	title        = {Spectral characteristics of {Common Iora} \textit{Aegithina} \textit{tiphia} vocalizations and their context-specific preferences},
	author       = {Divyapriya, C and Pramod, P},
	year         = {2019},
	journal      = {Current Science},
	publisher    = {JSTOR},
	volume       = {117},
	number       = {11},
	pages        = {1863--1871}
}

@misc{forrest2024cbc,
	title        = {Cbc (Coin-or Branch and Cut) Solver},
	author       = {Forrest, John and Lougee-Heimer, Robin},
	year         = {2024},
	url          = {https://github.com/coin-or/Cbc},
	note         = {COIN-OR Open-Source Mixed Integer Programming Solver}
}

@article{fraixedas2020state,
	title        = {A state-of-the-art review on birds as indicators of biodiversity: Advances, challenges, and future directions},
	author       = {Fraixedas, Sara and Lind{\'e}n, Andreas and Piha, Markus and Cabeza, Mar and Gregory, Richard and Lehikoinen, Aleksi},
	year         = {2020},
	journal      = {Ecological Indicators},
	publisher    = {Elsevier},
	volume       = {118},
	pages        = {106728}
}

@article{funosas2026global,
	title        = {A global assessment of {BirdNET} performance: Differences among continents, biomes, and species},
	author       = {Funosas, David and Sebasti{\'a}n-Gonz{\'a}lez, Esther and Morant, Jon and G{\'o}mez, Oscar H Mar{\'\i}n and Mendoza, Irene and Mohedano-Mu{\~n}oz, Miguel A and Santamar{\'\i}a, Eduardo and Bastianelli, Giulia and M{\'a}rquez-Rodr{\'\i}guez, Alba and Budka, Micha{\l} and others},
	year         = {2026},
	journal      = {Ecological Indicators},
	publisher    = {Elsevier},
	volume       = {182},
	pages        = {114550}
}

@misc{he2016deep,
	title        = {Deep Residual Learning for Image Recognition},
	author       = {He, Kaiming and Zhang, Xiangyu and Ren, Shaoqing and Sun, Jian},
	year         = {2016},
	howpublished = {In \emph{Proceedings of the IEEE Conference on Computer Vision and Pattern Recognition (CVPR)}, pp.\ 770--778, Las Vegas, NV. IEEE},
	doi          = {10.1109/CVPR.2016.90}
}

@misc{howard2019searching,
	title        = {Searching for {MobileNetV3}},
	author       = {Howard, Andrew and Sandler, Mark and Chu, Grace and Chen, Liang-Chieh and Chen, Bo and Tan, Mingxing and Wang, Weijun and Zhu, Yukun and Pang, Ruoming and Vasudevan, Vijay and others},
	year         = {2019},
	howpublished = {In \emph{Proceedings of the IEEE/CVF International Conference on Computer Vision}, pp.\ 1314--1324, Seoul, South Korea. IEEE}
}

@article{kahl2021birdnet,
	title        = {{BirdNET}: A deep learning solution for avian diversity monitoring},
	author       = {Kahl, Stefan and Wood, Connor M and Eibl, Maximilian and Klinck, Holger},
	year         = {2021},
	journal      = {Ecological Informatics},
	publisher    = {Elsevier},
	volume       = {61},
	pages        = {101236},
	doi          = {10.1016/j.ecoinf.2021.101236}
}

@misc{kahl2023overview,
	title        = {Overview of {BirdCLEF} 2023: Bird acoustic detection and identification},
	author       = {Kahl, Stefan and Vellinga, Willem-Pier and Denton, Samuel and Flinsenberg, Stefan and Fedorov, Roman and Klinck, Holger and Planque, Robert and Glotin, Herv{\'e}},
	year         = {2023},
	howpublished = {In \emph{Working Notes of CLEF 2023}, Vol.\ 3497, pp.\ 1--12, Thessaloniki, Greece. CEUR-WS}
}

@misc{lepage2026,
	title        = {Checklist of the birds of {M}alaysia},
	author       = {Lepage, Denis},
	year         = {2026},
	publisher    = {Avibase: The World Bird Database},
	url          = {https://avibase.bsc-eoc.org/},
	note         = {Accessed: 2026-03-10}
}

@misc{mybis2026,
	title        = {Birds of {M}alaysia},
	author       = {{Malaysia Biodiversity Information System (MyBIS)}},
	year         = {2026},
	url          = {https://www.mybis.gov.my/},
	note         = {Accessed: 2026-03-10}
}

@article{park2019specaugment,
	title        = {{SpecAugment}: A simple data augmentation method for automatic speech recognition},
	author       = {Park, Daniel S and Chan, William and Zhang, Yu and Chiu, Chung-Cheng and Zoph, Barret and Cubuk, Ekin D and Le, Quoc V},
	year         = {2019},
	journal      = {arXiv preprint arXiv:1904.08779}
}

@article{perez2023birdnet,
	title        = {{BirdNET}: applications, performance, pitfalls and future opportunities},
	author       = {P{\'e}rez-Granados, Cristian},
	year         = {2023},
	journal      = {Ibis},
	publisher    = {Wiley Online Library},
	volume       = {165},
	number       = {3},
	pages        = {1068--1075}
}

@article{puan2019influence,
	title        = {Influence of landscape matrix on urban bird abundance: evidence from {Malaysian} citizen science data},
	author       = {Puan, Chong Leong and Yeong, Kok Loong and Ong, Kang Woei and Fauzi, Muhd Izzat Ahmad and Yahya, Muhammad Syafiq and Khoo, Swee Seng},
	year         = {2019},
	journal      = {Journal of Asia-Pacific Biodiversity},
	publisher    = {Elsevier},
	volume       = {12},
	number       = {3},
	pages        = {369--375}
}

@book{rasmussen2012birds,
	title        = {Birds of {South Asia}: The {Ripley} Guide},
	author       = {Pamela C. Rasmussen and John C. Anderton},
	year         = {2012},
	publisher    = {Smithsonian Institution and Lynx Edicions},
	address      = {Washington, D.C. and Barcelona},
	isbn         = {978-84-96553-88-9},
	note         = {Two volumes; Volume 1: Field Guide, Volume 2: Attributes and Status. Notes complex, variable vocalisations and mimicry-like phrase diversity in Common Iora},
	edition      = {2nd}
}

@article{sebastian2025geographic,
	title        = {Geographic variation in acoustic signals in wildlife: A systematic review},
	author       = {Sebasti{\'a}n-Gonz{\'a}lez, Esther and P{\'e}rez-Granados, Cristian},
	year         = {2025},
	journal      = {Journal of Biogeography},
	publisher    = {John Wiley and Sons Inc},
	volume       = {52},
	number       = {6},
	pages        = {e15116},
	doi          = {10.1111/jbi.15116},
	issn         = {13652699}
}

@misc{sprengel2016audio,
	title        = {Audio Based Bird Species Identification using Deep Learning Techniques},
	author       = {Sprengel, Elias and Jaggi, Martin and Kilcher, Yannic and Hofmann, Thomas},
	year         = {2016},
	howpublished = {In \emph{CLEF 2016 Working Notes}, pp.\ 547--559, {\'E}vora, Portugal. CEUR-WS}
}

@article{stowell2022computational,
	title        = {Computational bioacoustics with deep learning: a review and roadmap},
	author       = {Stowell, Dan},
	year         = {2022},
	journal      = {PeerJ},
	publisher    = {PeerJ Inc.},
	volume       = {10},
	pages        = {e13152}
}

@misc{tan2019efficientnet,
	title        = {{EfficientNet}: Rethinking model scaling for convolutional neural networks},
	author       = {Tan, Mingxing and Le, Quoc},
	year         = {2019},
	howpublished = {In \emph{Proceedings of the 36th International Conference on Machine Learning (ICML)}, pp.\ 6105--6114, Long Beach, CA. PMLR}
}

@misc{zhang2018mixup,
	title        = {mixup: Beyond Empirical Risk Minimization},
	author       = {Zhang, Hongyi and Cisse, Moustapha and Dauphin, Yann N. and Lopez-Paz, David},
	year         = {2018},
	howpublished = {In \emph{6th International Conference on Learning Representations (ICLR)}, Vancouver, BC. OpenReview.net}
}

\end{document}